\documentclass[twocolumn]{webofc}

\usepackage[varg]{txfonts}   
\usepackage{hyperref}
\usepackage{url}
\hypersetup{colorlinks=true,citecolor=blue,urlcolor=blue,linkcolor=blue}
\newcommand{\be}{\begin{eqnarray}}
\newcommand{\ee}{\end{eqnarray}}
\newcommand{\ba}{\begin{array}}
\newcommand{\ea}{\end{array}}
\newcommand{\bi}{\begin{itemize}}
\newcommand{\ei}{\end{itemize}}

\begin{document}
\title{Bayesian Inference of fine-features of dense matter EOS from future high-precision data of neutron star radii}

\author{\firstname{Bao-An} \lastname{Li}\inst{1}\fnsep\thanks{\email{Bao-An.Li@etamu.edu}}
\and
        \firstname{Xavier} \lastname{Grundler}\inst{1}\and
        \firstname{Wen-Jie} \lastname{Xie}\inst{2,3}
        \and
        \firstname{Nai-Bo} \lastname{Zhang}\inst{4}
        \institute{Department of Physics and Astronomy, East Texas A$\&$M University, Commerce, TX 75429-3011, USA
\and Department of Physics, Yuncheng University, Yuncheng 044000, China
\and Guangxi Key Laboratory of Nuclear Physics and Nuclear Technology, Guangxi Normal University, Guilin 541004, China
\and School of Physics, Southeast University, Nanjing 211189, China}
}  
\abstract{
Future high-precision X-ray and gravitational wave observatories are expected to measure the radii of neutron stars (NSs) with an accuracy better than about 0.1 km. However, it remains unclear what particular aspects of the Equation of State (EOS) and to what precision they will be better constrained. Within a Bayesian framework using a meta-model EOS and mock high-precision NS data, the posterior probability distribution functions (PDFs) of NS matter EOS parameters for both hadronic and quark phases and the transition between them were recently studied. We report here a few highlights of these studies. }
\maketitle
\section{Introduction}
\label{intro}
\begin{figure}[ht]
\begin{center}
\resizebox{0.49\textwidth}{!}{
  \includegraphics{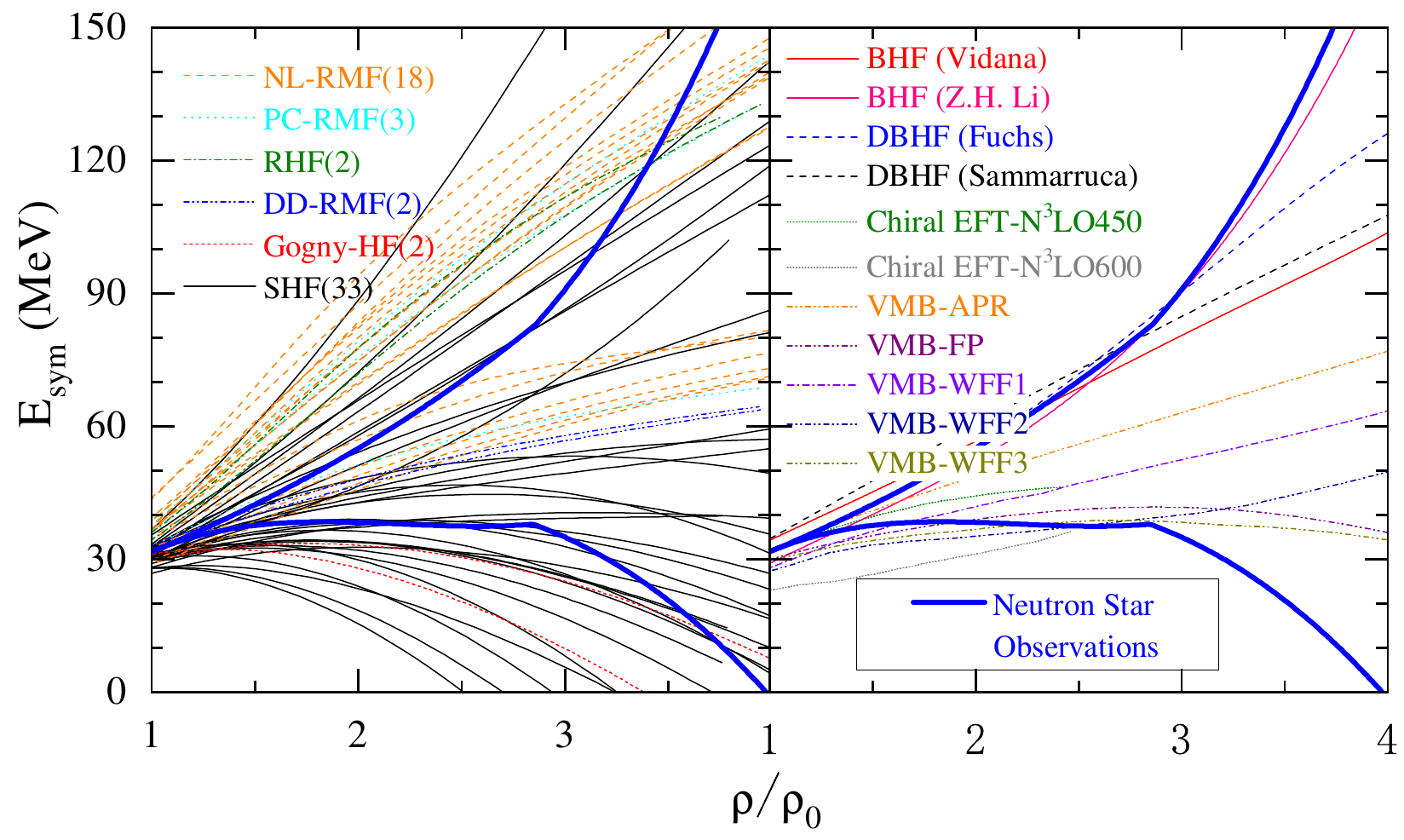}
  }
    \end{center}
    \vspace{-0.5cm}
    \caption{Left: 60 examples of predicted $E_{\rm{sym}}(\rho )$ using 6 classes of nuclear energy density functionals. Right: predictions using 11 microscopic and/or {\it ab initio} nuclear many-body theories \cite{ChenLW14}. In both panels, they are compared with the upper and lower boundaries of symmetry energy extracted from analyzing neutron star observations \cite{Zhang19apj,Xie19}. Taken from ref. \cite{Zhang19epj}.}\label{Esym-all}
\end{figure}

The current accuracy of measuring NS radii is about 1 km. For instance, LIGO/VIRGO inferred a radius of $R_{1.4}=11.9\pm 0.875$ km at 68\% confidence level for two canonical NSs involved in GW170817 \cite{LIGO18}. More recent NICER observations for NS radii generally have similar or larger errors, see, e.g., Refs. \cite{Riley19,Miller19}. These data have been used by the nuclear astrophysics community in extracting information about the nature and EOS of supradense NS matter in various ways, with many interesting results. In particular, they have helped us in better constraining the density dependence of nuclear symmetry energy $E_{\rm{sym}}(\rho )$ at densities around $(1-2)\rho_0$ where $\rho_0$ is the saturation density of symmetric nuclear matter, see, e.g., Ref.\ \cite{LCXZ2021} for a comprehensive review. As an example, shown in Fig. \ref{Esym-all} is a comparison of the constraints on $E_{\rm{sym}}(\rho )$ from NS observations (thick blue curves) with predictions of both phenomenological (left) and microscopic (right) nuclear many-body theories. Interestingly, while NS observations can already rule out many theoretical predictions around $\rho_0$, there is a large open window of observational constraints at $\rho \geq 2\rho_0$. More precise NS radius measurements may help close this open window. Moreover, to distinguish many possible EOSs and identify twin stars or strange stars, it is necessary to carry out the differential mass/radius measurement ${\rm d}M/{\rm d}R$ in the mass region $(1.2-2.0)$ M$_{\odot}$ \cite{Li24-PRD,Zhao20,Han22,Pro,Zhang25,Cai25,Xavier}. For this purpose, NS radii have to be measured much better than the current 1.0 km accuracy. 

It is very encouraging to note that high-precision NS radius measurements will be possible by using the next-generation X-ray pulse profile observatories and gravitational wave detectors. For instance, the enhanced X-ray Timing and Polarimetry mission (eXTP) \cite{AngLi25} to be launched around 2030 is designed to measure the radius of PSR 10740+6620 to about $\pm 6\%$ accuracy, while the Advanced Telescope for High Energy Astrophysics (NewATHENA) \cite{Ath} to be launched around 2037 can measure the radius of PSR 10740+6620 (NICER has measured its radius with a precision of about 10\% of its mean radius) to about $\pm 3\%$ accuracy. 
More excitingly, the third-generation gravitational-wave detectors \cite{Hild:2009ns,LIGOScientific:2020zkf}, e.g., Einstein Telescope \cite{Sathyaprakash:2012jk} and Cosmic Explorer \cite{Evans:2021gyd} may measure the radii at even higher precision. For example, based on several recent analyses and simulations, see, e.g., Refs. \cite{Chatziioannou:2021tdi,Pacilio:2021jmq,Bandopadhyay:2024zrr,Finstad:2022oni,Walker:2024loo}, the planned new gravitational wave facilities can measure the $R_{1.4}$ to a precision better than 2.0\%. 

Given the super-difficult work involved and the super-expensive investments needed to obtain the super-precise neutron star radius data, it is important to study now what new physics we can learn about the NS EOS from the expected high-precision data. For this purpose, we have recently performed Bayesian analyses using mock high-precision NS radius data and a meta-model for NS EOS \cite{Li24-PRD,Li25-APJ}. In the following, we summarize our key findings after introducing some necessary notations about NS EOSs. 

\section{Meta-model EOS for Bayesian analyses of NS observables}
First, we use the minimum model of NS matter consisting of neutrons, protons, electrons and muons in $\beta$ equilibrium to construct the hadronic part of the NS EOS. In this model, the most basic input to construct the NS matter EOS $P(\epsilon$), i.e., pressure $P$ versus energy density $\epsilon$, is the energy per nucleon $E(\rho,\delta )$ in asymmetric nucleonic matter of isospin asymmetry $\delta$ and density $\rho$. It can be written as
$E(\rho ,\delta )=E_0(\rho)+E_{\rm{sym}}(\rho )\delta ^{2} +\mathcal{O}(\delta^4)$
in terms of the symmetric nuclear matter (SNM) EOS $E_0(\rho)$ and nuclear symmetry energy $E_{\rm{sym}}(\rho )$. In Bayesian analyses, they can be parameterized respectively by
\begin{eqnarray}\label{E0para}
E_{0}(\rho)&=&E_0(\rho_0)+\frac{K_0}{2}(\frac{\rho-\rho_0}{3\rho_0})^2+\frac{J_0}{6}(\frac{\rho-\rho_0}{3\rho_0})^3\nonumber\\
E_{\rm{sym}}(\rho)&=&E_{\rm{sym}}(\rho_0)+L(\frac{\rho-\rho_0}{3\rho_0})\nonumber\\
&+&\frac{K_{\rm{sym}}}{2}(\frac{\rho-\rho_0}{3\rho_0})^2
+\frac{J_{\rm{sym}}}{6}(\frac{\rho-\rho_0}{3\rho_0})^3\label{Esympara}.
\end{eqnarray}
At the asymptotical limit of $\rho\rightarrow \rho_0$, the parameters involved will obtain their following physical meaning: 
(1) the $K_0$ and $J_0$ are the incompressibility and skewness of SNM, (2) the $E_{\rm{sym}}(\rho_0)$, $L$, $K_{\rm{sym}}$, and $J_{\rm{sym}}$ are the magnitude, slope, curvature, and skewness of symmetry energy at $\rho_0$. At abnormal densities, they are just parameters for first preparing the prior EOSs and then reconstructing the posterior EOS in Bayesian analyses.
This connection allows us to use empirical values of these parameters to set their prior range in Bayesian analyses, see, e.g., Refs. \cite{Zhang18,Xie20} for detailed discussions. 

\begin{figure}[h!]
\centering
\resizebox{0.5\textwidth}{!}{
\includegraphics[height=7.cm]{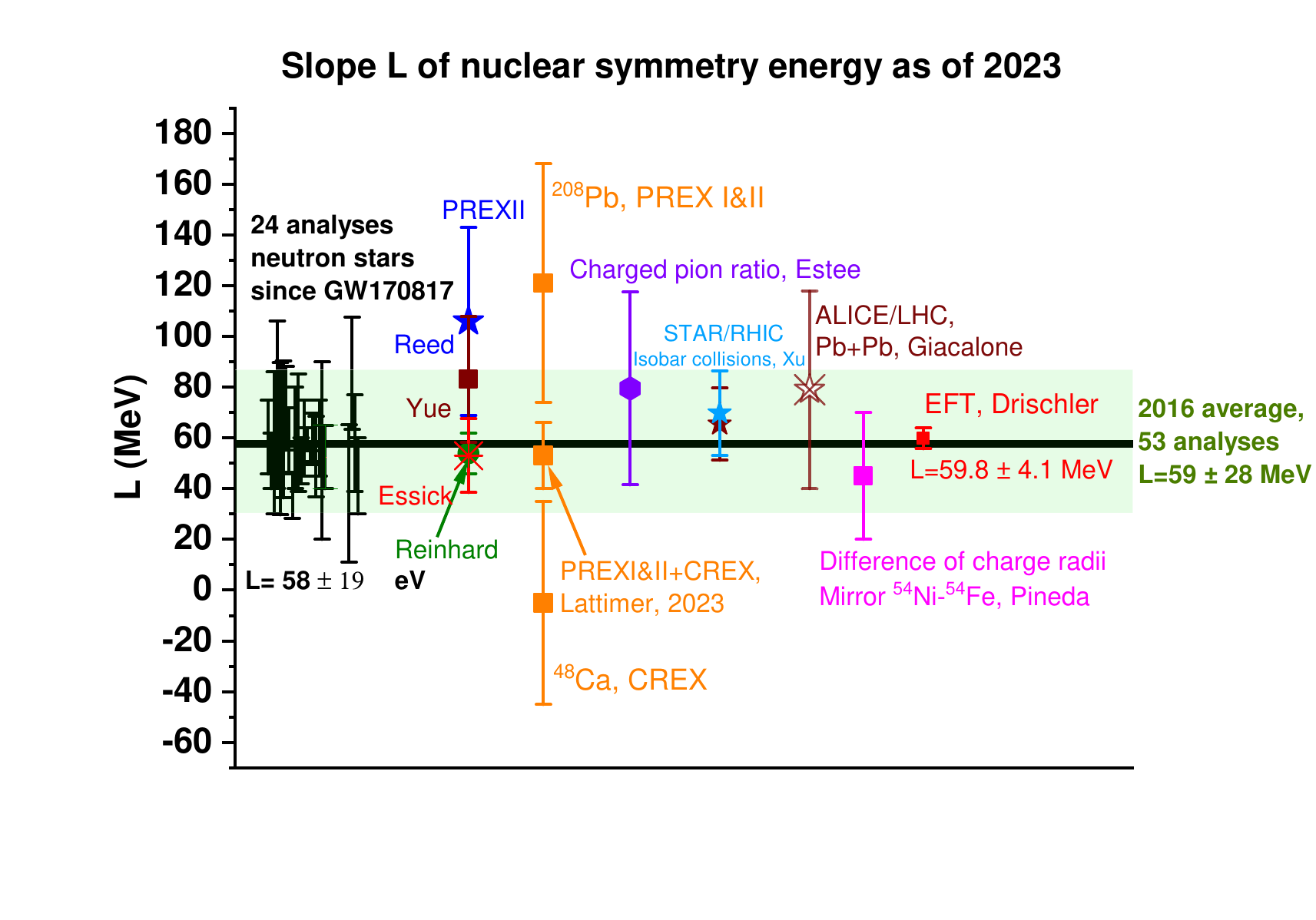}
}
\vspace{-1.2cm}
\caption{Updated constraints on the slope parameter $L$ of symmetry energy up to the year 2023. The shaded band around $L=59\pm 28$ MeV is from the 2016 survey. The 24 black bars represent 24 independent analyses of NS observables conducted within two years after GW170817. They gave a mean value of $L=58\pm 19$ MeV. Taken from Ref.\,\cite{Li25B}}\label{L2023}
\end{figure}

\begin{figure}[ht]
\begin{center}
\hspace*{-0.6cm}
\resizebox{0.5\textwidth}{!}{
  \includegraphics[width=7cm]{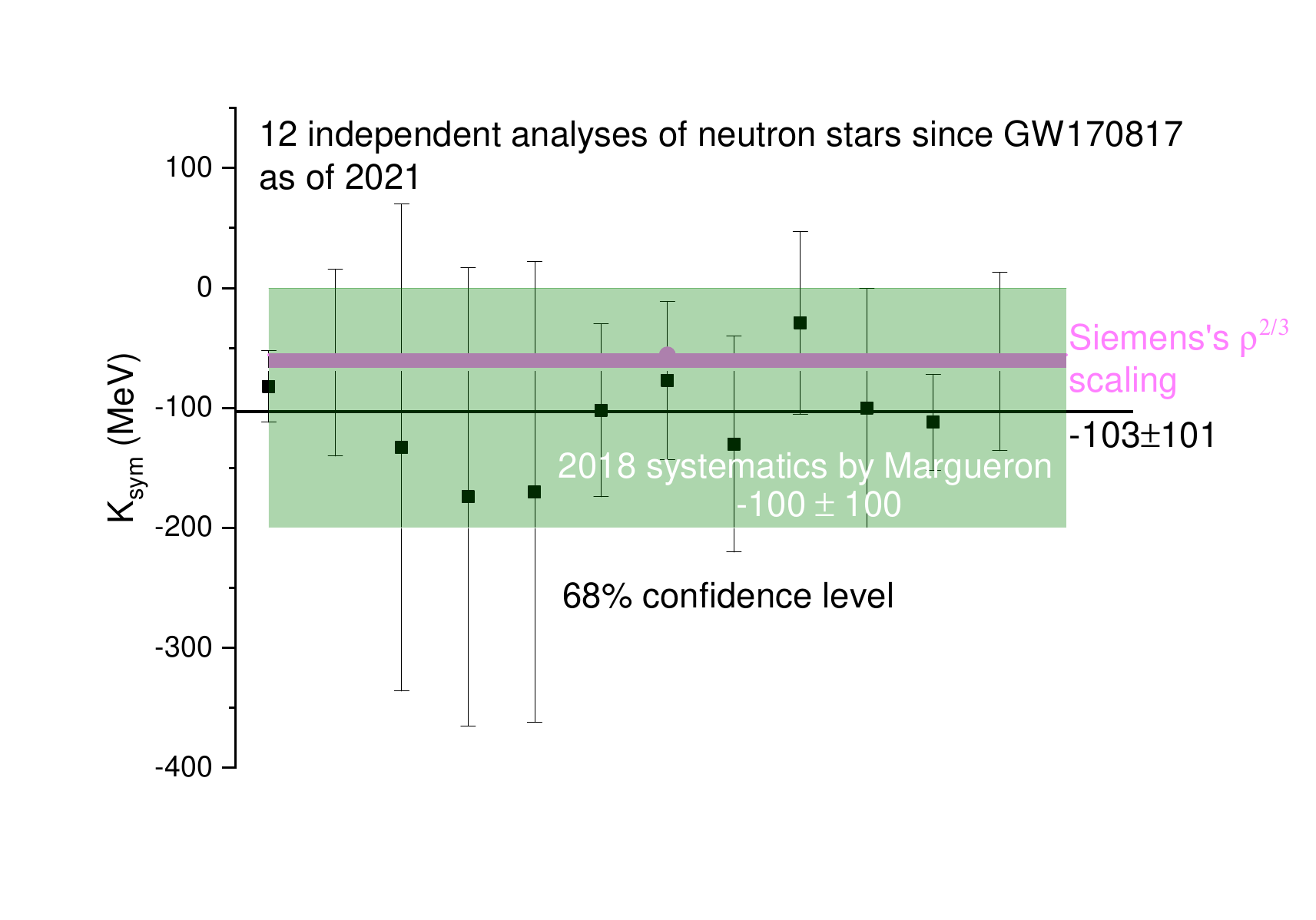}
  }
\vspace{-1.cm}
\caption{The curvature parameter $K_{\rm{sym}}$ at 68\% confidence level from 12 independent analyses of NSs (listed in Ref. \cite{LCXZ2021}). They gave a mean value of $K_{\rm{sym}}=-103\pm 101$ MeV. The green band within $-100\pm 100$ MeV is from the 2018 survey by Margueron et al.~\cite{Margueron18}. The pink band within $-(56-68)$ MeV is from Siemens' $\rho^{2/3}$ scaling of symmetry energy \cite{Phil}. Taken from Ref. \cite{Li25B}.}\label{Ksym}
\end{center}
\end{figure}

Among all the EOS parameters for hadronic matter, the $E_0(\rho_0)$, $K_0$, and $E_{\rm{sym}}(\rho_0)$ are reasonably well determined. While significant progress especially since GW170817 has been made in determining $L$ and $K_{\rm{sym}}$ as illustrated in Fig. \ref{L2023} and Fig. \ref{Ksym} from a recent survey \cite{Li25B}, very little is known about $J_{\rm{sym}}$ which characterizes the behavior of $E_{\rm{sym}}(\rho)$ at densities above about $3\rho_0$. On the other hand, there are some constraints on $J_0$ from analyzing collective flow in relativistic heavy-ion collisions \cite{Xie-JPG} and masses of NSs \cite{Zhang-M217}.

To investigate if the precision of NS radius measurement has any effect on inferring quark matter (QM) and/or hadron-quark transition properties, we also consider hybrid stars described by a hadron-quark first-order phase transition using the so-called constant sound speed (CSS) model \cite{Alford:2013aca}
\begin{equation}\label{css}
    \varepsilon(p)= \begin{cases}\varepsilon_{\mathrm{HM}}(p) & p<p_{t} \\ \varepsilon_{\mathrm{HM}}\left(p_{t}\right)+\Delta \varepsilon+c_{\mathrm{qm}}^{-2}\left(p-p_{t}\right) & p>p_{t}\end{cases}
\end{equation}
where $\varepsilon_{\rm HM} (p)$ is the EOS of hadronic matter (HM) described above, and $p_t$ is the pressure at the phase transition. The parameters we use are the transition density, $\rho_t/\rho_0$, which determines the transition pressure, $p_t$, the energy density discontinuity at the hadron-quark interface, $\Delta \varepsilon / \varepsilon_t$, which describes the strength of the phase transition, and the speed of sound squared in QM, $c_{\rm qm}^2$, which describes the stiffness of QM. By initializing randomly all nine parameters in their prior ranges, we can mimic most or all other proposed NS EOS in the literature. This model provides a simple but useful reference for more complicated models for the possible formation of QM in NSs \cite{Xavier}.

\begin{figure}[!ht]
\begin{center}
\includegraphics[width=0.49\textwidth]{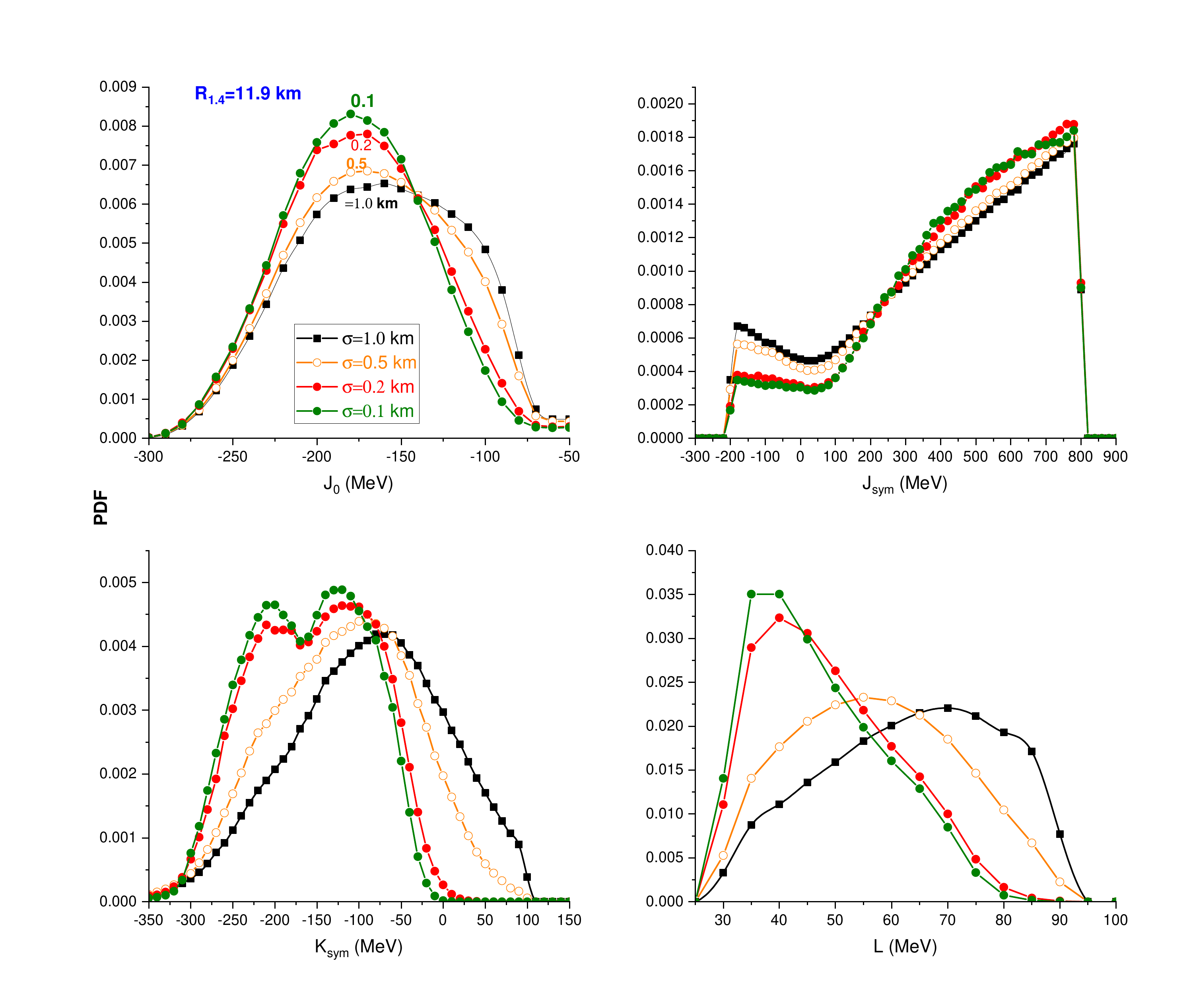}
\end{center}
\vspace{-0.8cm}
 \caption{Posterior PDFs of EOS parameters with $R_{1.4}=11.9$ km and a precision of $\Delta R=1.0, 0.5, 0.2$, and $0.1$ km, respectively. The figure is taken from Ref. \cite{Li24-PRD}. }\label{R14PDF}
 \end{figure}\section{Bayesian inferences of EOS parameters from high-precision NS radius data}
Shown in Fig. \ref{R14PDF} is a comparison of the posterior PDFs of $J_0$, $J_{\rm{sym}}$, $K_{\rm{sym}}$, and $L$ from Bayesian analyses using the minimum model of NSs and the fiducial radius data of $R_{1.4}=11.9$ km for a canonical NS with an imagined precision of $\Delta R=1.0, 0.5, 0.2$, and $0.1$ km, respectively. Several interesting observations are worth emphasizing. 

First, while the most probable value of the skewness $J_0$ measuring the stiffness of high-density SNM remains about the same, its precision gets appreciably improved as the precision $\Delta R$ changes from 1.0 km to 0.1 km, indicating the potential of better constraining the EOS of high-density SNM. 

Second, the PDFs of $L$ and $K_{\rm{sym}}$ have significant changes. In particular, the most probable $L$ shifts to smaller values and the PDF($K_{\rm{sym}}$) starts to show two peaks as the precision improves. It is known that these two parameters are most important for determining the radii of canonical NSs \cite{Jake}. The dual peaks in the PDF($K_{\rm{sym}}$) is due to the correlations between $L-K_{\rm{sym}}$ and $K_{\rm{sym}}-J_{\rm{sym}}$ \cite{Li24-PRD}. It is not surprising that as the precision of probes improves, fine structures of the object get revealed, as seen frequently in many sub-fields of science. On the other hand, the PDF($J_{\rm{sym}}$) peaks at the upper boundary of its uniform prior. These findings are qualitatively expected \cite{Xie19,Xie20}. It is known that the radii of canonical neutron stars are determined by the pressure around $2\rho_0$ to which the symmetry energy makes a major contribution. Thus, using the same mean radius $R_{1.4}=11.9$ km with better precision from 1.0 to 0.1 km will lead mainly to a more precise inference of nuclear symmetry energy around $2\rho_0$ characterized by $L$ and $K_{\rm{sym}}$. It will not improve much about the EOS at significantly higher densities characterized by $J_0$ and $J_{\rm{sym}}$. 

Third, the response of the PDFs of symmetry energy parameters to the variation of the precision $\Delta R$ is rather asymmetric. This is rather different from the response of SNM skewness $J_0$.
It is seen that the PDF$(J_0)$ mostly narrows its width symmetrically around approximately the same most probable value. It is known that the $J_0$ is mainly determined by the maximum mass and causality, while the radii are mainly controlled by the symmetry energy. Thus, the PDF$(J_0)$ is not affected much by the variation of $\Delta R$. On the other hand, a small change in radius can cause a significant change in the density profile or average density of NSs. Since the Tolman–Oppenheimer–Volkoff equations governing NS structures are highly nonlinear, the mapping between the mass-radius sequence and the EOS is rather nonlinear. 

\begin{figure*}[ht]
\centering
 \resizebox{0.8\textwidth}{!}{
  \includegraphics[trim={5mm, 90mm, 5mm, 7mm}, clip, width=16cm]{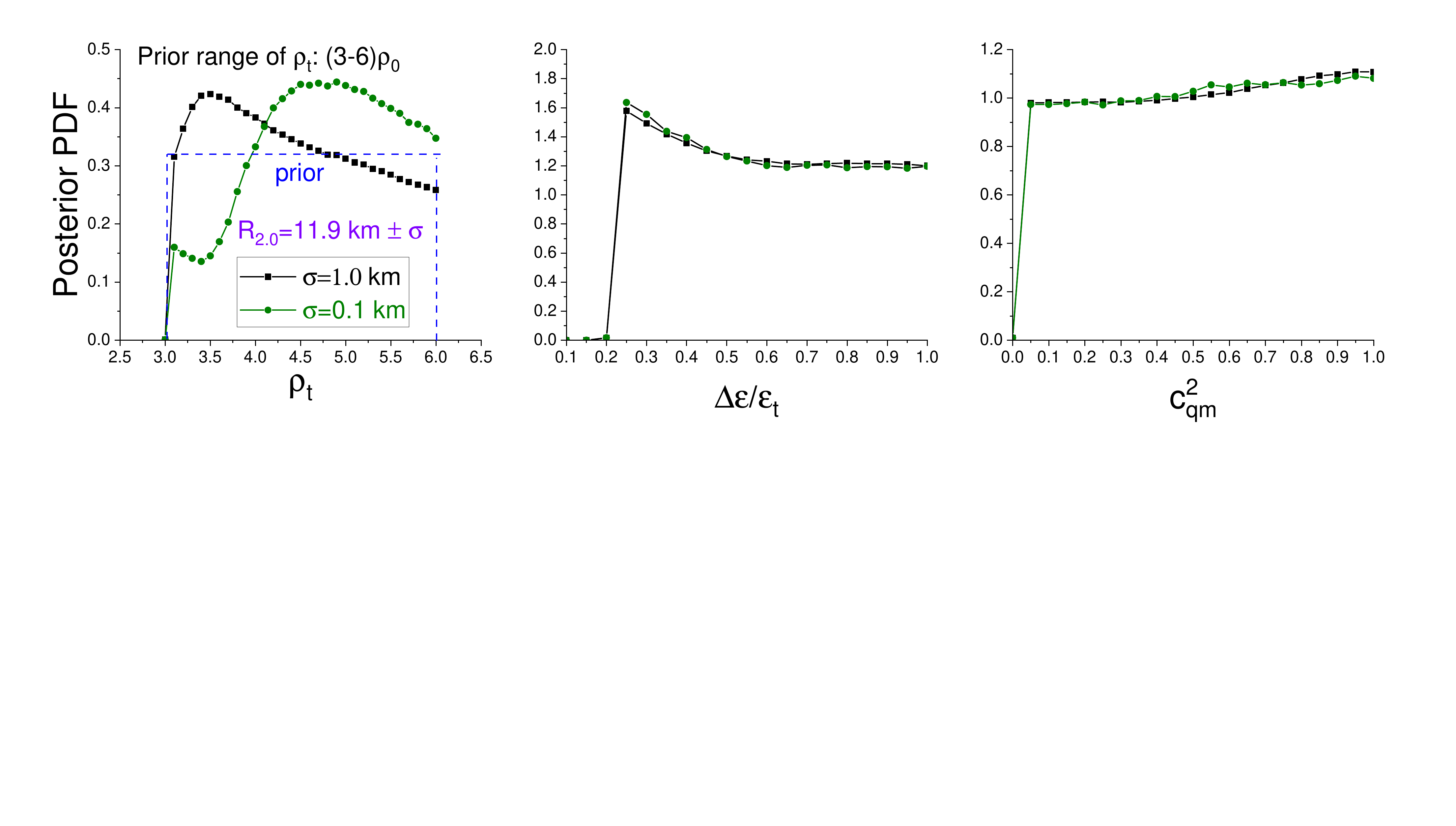}
  }
\setlength{\abovecaptionskip}{-2.cm}
\caption{Posterior PDFs of the three quark matter EOS parameters inferred from $R_{2.0}=11.9$ km data using $(3.0-6.0)\rho_0$ as the prior range for $\rho_t$ and precision $\sigma=1.0$ and $0.1$ km, respectively. Taken from Ref.~\cite{Li25-APJ}}.\label{QM36R20}
\end{figure*}

We now turn to the effects of precision $\sigma=\Delta R$ on inferring the EOS of QM and properties of hadron-quark first-order phase transition within the CSS model. Shown in Fig. \ref{QM36R20} are the posterior PDFs of the three quark matter EOS parameters inferred from $R_{2.0}=11.9$ km with a precision of $\sigma=1.0$ and $0.1$ km, respectively. In this example, the prior range for the hadron-quark transition density is set $3.0\leq \rho_t/\rho_0\leq 6.0$. The reason we now use the mock radius data for the radii of NSs with 2.0M$_{\odot}$ instead of canonical NSs with 1.4M$_{\odot}$ is because the radii of more massive NSs are less affected by the still uncertain crust EOSs, thus more reliable for probing the properties of QM and hadron-quark phase transition.  

It is interesting to see that the PDF and the most probable value of $\rho_t$ are very sensitive to the precision of NS radius measurement. This is understandable because the average density $\rho_a$ of an NS scales with $M/R^3$; a small variation of its radius can lead to a big change in its $\rho_a$ and the entire density profile. Moreover, as we mentioned earlier, the correspondence between the NS mass-radius relation and the underlying EOS, especially in the high-density region, is highly nonlinear because of the highly nonlinear nature of the TOV equations. 

It is also very interesting to see that the precision $\sigma$ has essentially no effect on the posterior PDFs of QM properties quantified by $\Delta\epsilon$ and $C^2_{\rm{qm}}$. Moreover, the PDF($C^2_{\rm{qm}}$) is rather flat in its whole range. It indicates that the NS radius data, regardless of its precision, does not constrain the quark matter stiffness. This is physical, although it may sound very disappointing to some people. In fact, it has been well known that the radii of canonical NSs are determined by the pressure at densities around $2\rho_0$ \citep{Lat}. For massive NSs, the relevant density is expected to be higher \citep{Cai25B}. But with the most probable hadron-quark transition density $\rho_t$ as high as $3.5\rho_0$ with $\sigma=1.0$ km and $4.7\rho_0$ with $\sigma=0.1$ km, the QM stiffness is not expected to affect the radii even for massive NSs. 

\section{Conclusions}
High-precision NS radius data from the forthcoming new measurements with the next-generation X-ray and gravitational wave detectors
will help constrain more stringently the high-density behavior of nuclear symmetry energy and the hadron-quark transition density.
But they will not affect much the inference of the quark matter EOS in NS cores.\\

\noindent{\bf Acknowledgement:} BAL and XG were supported in part by the U.S. Department of Energy, Office of Science, under Award Number DE-SC0013702, the CUSTIPEN (China-U.S. Theory Institute for Physics with Exotic Nuclei) under the US Department of Energy Grant No. DE-SC0009971. WJX was supported in part by the Shanxi Provincial Foundation for Returned Overseas Scholars under Grant No 20220037, the Natural Science Foundation of Shanxi Province under Grant No 20210302123085, the Open Project of Guangxi Key Laboratory of Nuclear Physics and Nuclear Technology, No. NLK2023-03 and the Central Government Guidance Funds for Local Scientific and Technological Development, China (No. Guike ZY22096024). NBZ is supported in part by the National Natural Science Foundation of China under Grant No. 12375120, the Zhishan Young Scholar of Southeast University under Grant No. 2242024RCB0013.
%
%

\newcommand{\apjl}{Astrophys. J. Lett.\ }
\newcommand{\apj}{Astrophys. J. \ }
\newcommand{\prl}{Phys. Rev. Lett.\ }
\newcommand{\prc}{Phys. Rev. C\ }
\newcommand{\prd}{Phys. Rev. D\ }
\newcommand{\mnras}{Mon. Not. R. Astron. Soc.\ }
\newcommand{\aap}{Astron. Astrophys.\ }
\newcommand{\nphysa}{Nucl. Phys. A\ }
\newcommand{\physrep}{Phys. Rep.\ }
\newcommand{\nat}{Nature\ }



\end{document}